\documentclass[a4paper]{jpconf}
\usepackage{graphicx}
\def\be{\begin{equation}}
\def\ee{\end{equation}}
\def\bea{\begin{eqnarray}}
\def\eea{\end{eqnarray}}

\newcommand{\mbf}[1]{\mathbf{#1}}

\begin{document}
\title{Neutrino mass, mixing and discrete symmetries
}

\author{Alexei Y. Smirnov~\footnote{Talk given at the Symposium  
Discrete 2012,   IST, Lisboa, Portugal, December 3 - 7, 2012.}}

\address{International Centre for Theoretical Physics, Strada Costiera 
11, 34014 Trieste, Italy}

\ead{smirnov@ictp.it}

\begin{abstract}
Status of the discrete symmetry approach to explanation 
of the lepton masses and mixing is 
summarized  in view of recent experimental results, 
in particular,  establishing relatively large 1-3 mixing. 
The lepton mixing can originate from breaking of discrete flavor symmetry $G_f$ to different residual 
symmetries $G_{\ell}$ and $G_\nu$ in the charged lepton and neutrino  sectors. 
In this framework the {\it symmetry group condition} has been derived which allows  
to get relations between the lepton mixing elements 
immediately without explicit model building. 
The condition has been applied to different residual neutrino symmetries 
$G_\nu$. For  generic (mass independent) 
$G_\nu = {\bf Z}_2$ 
the condition leads to two relations between the mixing parameters and fixes 
one column of the mixing matrix.  
In the case of  $ G_\nu = {\bf Z}_2 \times {\bf Z}_2$ the condition  
fixes the mixing matrix completely. The non-generic (mass spectrum dependent)  
$G_\nu$  lead to relations which include mixing angles, neutrino masses and Majorana 
phases.  The symmetries $G_{\ell}$, $G_\nu$, $G_f$ are identified 
which lead to the experimentally observed values of the mixing angles and allow to predict 
the CP phase. 


\end{abstract}

\section{Introduction}

A possibility to use the discrete flavor symmetries  
for understanding fermion masses and  mixing had been 
proposed long time ago~\cite{hist}.  
To a large extend the  recent developments of this approach~\cite{reviews} 
was motivated by the tri-bimaximal (TBM)  mixing 
\cite{tbm}. 
The special values of elements of the TBM mixing matrix, 
\be 
|U_{e3}|^2 = \sin^2 \theta_{13} = 0,~~~~ 
|U_{\mu 3}|^2 =  |U_{\tau 3}|^2  = \sin^2 \theta_{23} = \frac{1}{2}, ~~~~
|U_{e2}|^2 = \sin^2 \theta_{12} = \frac{1}{3},  
\label{tbmparam}
\ee
indicated possible geometric and group-theoretical
origins of the mixing.  The values in (\ref{tbmparam})  also 
implied that there is no relation between mixing and 
masses (mass ratios). 
In this connection the most appealing framework is the one in which mixing originates 
from different ways of the flavor symmetry, $G_f$,   breaking in the charged 
lepton and neutrino  Yukawa sectors~\cite{framework}:  
\be 
G_f \rightarrow  {\rm breaking} \rightarrow \left\{ 
\begin{array}{lll}
 G_\nu  &  {\rm neutrinos}\\
 G_l &  {\rm charged~~ leptons} 
\end{array}
\right. . 
\label{framework}
\ee
The {\it residual} symmetries $G_\nu$ and $G_l$
of  the neutrino  and charged lepton mass matrices are different. 
$G_\nu$ and $G_l$ should be generic symmetries which 
exist for arbitrary values of masses. 
In this case appearance of TBM is, indeed,  maximally controlled by symmetry.  

Realizations of this framework in specific models are, however, rather complicated and 
not convincing.  One  should construct the Lagrangian  
invariant under symmetry $G_f$  
with certain field content and  assignment of charges.  
Additional auxiliary symmetries are needed to forbid some interactions. 
Further model building is needed to achieve required vacuum alignment.   
Essentially, two different scalar sector should be introduced 
for spontaneous symmetry breaking in the neutrino and charged lepton sectors.  
Finally, after spontaneous symmetry breaking the mass matrices  
with certain residual symmetries emerge. 
Accidental symmetries and relations between mixing elements 
may show up in models   
which are not related to the original flavor symmetry.  
Complicated structure of models, many {\it ad hoc} parameters and new fields, 
additional symmetries, difficulty to include quarks, {\it etc.} 
cast doubt upon whole the approach \cite{my}.

Recent measurements of the neutrino oscillation parameters,  
and especially discovery of rather large 
1-3 mixing~\cite{T2K13}, \cite{daya}, \cite{RENO}, \cite{DC}, 
\cite{minos13}, contrary to (\ref{tbmparam}),  have further disfavored TBM 
and the discrete symmetries behind.  
The TBM can be  accidental and whole discrete 
flavor symmetry approach -- phenomenologically irrelevant. 

Further developments along this line 
were in two directions: 
(i) introduction of  large corrections to TBM 
to reproduce results of measurements \cite{corr},  (ii)
modification of symmetries 
in such a way that they are  consistent with nonzero 1-3 
mixing \cite{modific}. In some situations these two lines coincide giving the same results. 
With this the symmetry effects become rather hidden. 

In this connection in the framework of residual symmetries (\ref{framework}) 
a formalism has been developed 
\cite{dani1}, \cite{dani2},  \cite{dani3} which allows to obtain consequences of flavor symmetries 
for mass and mixing  without model building. The formalism  
allows to explore mixing patterns for wide class of different 
residual and covering symmetries. Inversely, it can be used to  perform the ``symmetry building'' 
for a required mixing pattern.   
It helps to understand various features of mixing that
are related to symmetries. 

Symmetries and their consequences are the nicest part of the program, 
model building is the ugly one. 
In this review we  will focus on the former: on the     
model independent part.
The formalism 
\cite{dani1, dani2} will be explained in details and 
various related issues will be clarified. The applications 
to different residual neutrino symmetries will be described.   
Still one should remember that the problems with model building 
remain.

The paper 
is organized as follows.  
In Sec.~2 we summarize the relevant information on 
neutrino masses and mixing and its possible 
implications for discrete flavor symmetries. 
Next possible step in developments of the field  will be outlined.  
In Sec.~3 the 
symmetry group conditions are derived which lead to relations 
between the lepton mixing elements without  model building. 
Applications of these symmetry group conditions to 
different residual neutrino symmetries are described in Sec. 4.     
Discussion and conclusions are presented in Sec.~5. 

\section{Masses and mixing: status and implications}


\subsection{Data and Observations.}

Results and observations from global fits \cite{fogli}, \cite{concha},  
\cite{valle} relevant for 
the discrete flavor symmetries can be summarized in the following way. 

1). The 1-3 mixing deviates substantially from zero: 
\be
\sin^2 \theta_{13} = 0.022 - 0.024. 
\label{eq:obs13}
\ee

2). There are indications of significant deviation of the 2-3 mixing from maximal: 
$\sin^2 \theta_{23} = 0.38 - 0.42$, 
which can be characterized by 
\be
d_{23} \equiv  0.5 \cos 2 \theta_{23} = 0.08 - 0.12.
\label{eq:obs23}
\ee 
The result follows from (i) the atmospheric neutrino data 
(included in the  global fit); (ii) from MINOS direct measurements~\cite{minos23}:  
$\sin^2 2\theta_{23} = 0.950^{+ 0.035}_{-0.036}$ ($1\sigma$),  
and (iii) from comparison of the T2K result \cite{T2K13} on 
$\nu_{\mu} \rightarrow \nu_{e}$ oscillations 
($ \propto \sin^2 \theta_{23} \sin^2 2\theta_{13}$) and the reactor 
disappearance results ($\propto \sin^2 2\theta_{13}$). 
The atmospheric neutrino and T2K data being sensitive to $\sin^2 \theta_{23}$ 
favor the first quadrant,  $\theta_{23} < \pi/4$,  at least for the normal mass hierarchy (NH). 
In the case inverted hierarchy (IH) different global fits give different results.  

3). With increase of accuracy the deviation of the 1-2 mixing 
from the TBM value 1/3 becomes stronger and more significant statistically:
\be
\sin^2 \theta_{12} = 0.30 - 0.31. 
\ee

4). First glimpses on the CP-violating phase show up  in 
different global fits:  $\delta = 180^{\circ}$ from \cite{fogli}, 
$\delta \sim 300^{\circ}$ for both hierarchies from \cite{concha},  
and  $\delta = 145^{\circ}$ (NH) $\delta = 0$ (IH) from \cite{valle}. 
Being at $1\sigma$ level or even lower  
all these are statistically insignificant.  
In some analyses one can trace the 
physical effects which favor certain value 
$\delta$,  in other cases that can be artifact of the global 
fit without any clear reason behind or just fluctuation.  

The deviation of  mixing from TBM  
can be characterized by 
the values of elements of the third column of the PMNS matrix,  
$U_{\alpha 3} \equiv \{U_{e3}, U_{\mu3}, U_{\tau 3} \}$:  
\be
U_{\alpha 3}^{TBM} = \{0,~ 0.71,~ 0.71\}, ~~~{\rm whereas}~~~
U_{\alpha 3}^{exp} = \{0.15,~ 0.62, ~0.77 \} ,   
\ee
and for the ``numerology bookkeeping'': 
$U_{\tau 3} - U_{\mu3} \approx U_{e3}$. 

\subsection{The 1-3 mixing: relations and implications.} 

The same value of 1-3 mixing can be 
connected to other observables in various ways  
which have different theoretical implications. 

{\it (i) ``Naturalness'':} 
\be
\sin^2 \theta_{13} = A
\frac{\Delta m_{21}^2}{\Delta m_{32}^2},
\label{eq:natur}
\ee
where  $A = 0.78$ 
($\sim 1 - \sin\theta_C$) for the best fit value of $\theta_{13}$. 
This relation
follows from ``naturalness'' of mass matrix \cite{akhm}: 
the fact that there are two large mixings connecting neighboring
generations, and  from the following  two assumptions: (i)
normal mass hierarchy, (ii) absence of fine tuning between different elements
of the mass matrix (e.g.,  $|m_{e\mu} -  m_{e \tau}| \sim |m_{e\mu}|$).

{\it (ii) ``QLC'':} 
\be
\sin \theta_{13} \approx \frac{1}{\sqrt{2}} \sin \theta_C
(1 - V_{cb} \cos \delta) - V_{ub} \approx \frac{\theta_C}{\sqrt{2}}, 
\label{eq:13qlc}
\ee
where $\theta_C$ is the Cabibbo angle \cite{qlc}.
Varying $\delta$ we have from (\ref{eq:13qlc}) 
$\sin^2 \theta_{13} = 0.025 \pm 0.003$
in agreement with observations (\ref{eq:obs13}, \ref{eq:obs23}). 
Approximately,  $\sin^2 \theta_{13} \approx \frac{1}{2} \sin^2 \theta_C$. 
This prediction is essentially result of permutation 
of the matrices of the maximal mixing 2-3 rotation and  the 1-2 rotation on
the Cabibbo angle:
$$
U_{12}(\theta_C) U_{23}\left(\frac{\pi}{4}\right).  
$$
Here $U_{12}(\theta_C)$ can follow from the charged leptons, 
whereas 
$U_{23}\left(\frac{\pi}{4}\right)$ -- from neutrinos. 
The former implies a kind of quark-lepton symmetry, 
or unification,  or common horizontal symmetry  
which lead to equality of the 1-2 mixings in the 
quark and lepton sectors. The second  
($\pi/4$) rotation  comes from the neutrino sector 
and can be part of the bi-maximal~\cite{qlc} 
or tri-bimaximal~\cite{king} mixings.  
The latter case is referred as the tribimaximal-Cabibbo mixing.
The permutation  is needed to reduce the mixing matrix to the standard form.  
The relation  was first realized at the purely phenomenological 
level~\cite{giunti} and then in the Quark-Lepton Complementarity (QLC)
framework~\cite{qlc}.  
Appearance of the bi-maximal or tri-bimaximal  mixings 
can be a consequence of discrete flavor symmetries. 

{\it (iii) The $\nu_\mu - \nu_\tau$ symmetry breaking}:  
\be
\sin^2 \theta_{13} = \frac{1}{2} \cos^2 2 \theta_{23}
= 2d_{23}^2 \approx 0.022
\label{eq:mutau}
\ee
in perfect agreement with measurements (\ref{eq:obs13},\ref{eq:obs23}). 
The relation (\ref{eq:mutau})  may follow from ``universal'' violation of the 
$\nu_\mu - \nu_\tau$ symmetry. (In the 
case of exact symmetry $\sin^2 \theta_{13} = d_{23} = 0$.) 

{\it (iv) The self-complementarity} \cite{selfcomp}: 
\be
\theta_{12} +  \theta_{13}  = \theta_{23}.
\label{eq:self}
\ee
It  is purely leptonic relation 
which is also reproduced by QLC.
 
{\it (v). ``Anarchy''}:  
\be
\sin^2 \theta_{13} > 0.025 . 
\ee
The inequality is realized at 
$1\sigma$ level in  the anarchy approach,  
in which values of mixing angles appear as random numbers \cite{anarchy}.

{\it (vi) ``Quark-lepton universality'':}  
$$
\theta_{13} \approx \frac{1}{2}\theta_{12}\theta_{23}.  
$$
This equality 
is similar to the one in the 
quark sector: $V_{ub} = \frac{1}{2} V_{us} V_{cb}$.
It may indicate that the mixing patterns of quarks and leptons 
are organized in the same way but with certain rescaling. 
That would testify for 
(i) a kind of Fritzsch ansatz for mass matrices;  
(ii)  normal mass hierarchy; 
(iii) relation between masses and mixing;  
(iv) flavor ordering  or alignment in the mass matrix.  
The later means that values of elements of the 
neutrino mass matrix in the flavor basis gradually decrease from  
$m_{\tau\tau}$  to $m_{ee}$.  In fact, due to deviation of the 2-3 mixing 
from maximal  sharp difference of the elements of the 
dominant  $\mu \tau$-block  and 
the sub-dominant  $e-$line is washed out.   
It seems that neutrino mass matrix is rescaling of the  
charged fermion mass matrices with  weaker hierarchy of the elements.  
This can originate from power dependence of elements 
on large expansion parameter   $\lambda  \sim 0.7 -  0.8$.  
It looks like another complementarity:  
$\lambda_\nu  = 1 - \lambda_q \approx \theta_C$. 
The power dependence could testify for a kind of 
Froggatt-Nielsen mechanism.

No one of these results explicitly testifies for discrete 
flavor symmetries, although, as we will see later, 
consequences of symmetry can be hidden  
in rather complicated relations between mixing parameters.  
It seems that relations $(i)$, $(iv)$ and $(vi)$ have no connection to  
discrete symmetries at all, the relation  $(ii)$ may require 
the symmetry to explain the maximal 2-3 rotation from neutrino sector, 
finally $(iii)$ may or may not be a consequence of symmetry.

Existence of sterile neutrinos with  the eV-scale mass 
can strongly affect our considerations. 
Their mixing with active neutrinos is not a small perturbation of the $3\nu-$ pattern: 
all theoretical constructions should be reconsidered 
unless further complications and fine tunings are introduced
\cite{rodejohan}.

\subsection{Race for the mass hierarchy.}

Future developments in the field will be related 
(apart from better measurements of known parameters) to 

\begin{itemize}

\item
clarification of the situation with sterile neutrinos; 

\item
checks of the claim of the neutrinoless 
double beta decay observation
in the Heidelberg-Moscow experiment;  
and 

\item
establishing neutrino mass hierarchy. 

\end{itemize}
All this will have serious impact on our understanding 
of mixings and masses and  
relevance of the discrete flavor symmetries.  

As far as the mass hierarchy is concerned, 
in  the $2\nu$ approximation changing the hierarchy means 
flip of the sign: $\Delta m_{31}^2 \rightarrow - \Delta m_{31}^2$. 
The discrete symmetry transformation can be introduced 
which makes this flip.  
The case of inverted hierarchy is the  special since it  implies 
strong degeneracy of the two mass states 
$\nu_1$ and $\nu_2$: $\Delta m/m \sim \Delta m_{21}^2/2 
\Delta m_{31}^2 \sim 1.6 \cdot 10^{-2}$. This strong degeneracy  can be a consequence of 
certain discrete symmetry. However it  is not accompanied by maximal mixing, which leads to certain 
tension. 

A number of proposals are participating in the ``race'' for the mass hierarchy. 
Many are related to influence of matter effects on the 1-3 mixing:  
In uniform or nearly uniform medium the resonance enhancement of oscillations driven 
by the 1-3 mixing is in the 
neutrino channel for NH and in the antineutrino channel in the case of IH. 
To some extend this can be explored using the beam experiments, 
e.g., in NOvA~\cite{nova}. 
The measured relatively large 1-3 mixing ensures that in supernova the 1-3 level crossing 
(H resonance) is highly adiabatic. Consequently,  
conversion effects in the cases of NH and IH are strongly different 
\cite{dighe}. That determines also the 
oscillation effects of SN neutrinos in the matter of the Earth. 
Observation of the oscillation effect in the antineutrino channel 
would be the proof of NH \cite{cecilia2}. 

Huge atmospheric neutrino detectors 
with low  ($\sim 1$ GeV) energy threshold such as 
PINGU~\cite{pingu} and ORCA have good chance to be the first 
and the cheapest ones \cite{ARS}. These experiments 
have also good potential to measure precisely the 2-3 mixing and mass splitting.

\section{Discrete symmetries and patterns of lepton mixing.}
We assume  that neutrinos are Majorana particles, 
and that the discrete flavor symmetry can be embedded into $SU(3)$. 
The latter requires that determinants of the transformation 
matrices should be 1.

\subsection{Residual symmetries and transformations. }

In what follows we will consider the framework of the residual symmetries~\cite{framework} 
in details. Let $S_\nu$  and $T$ be the transformations 
of the neutrino,  $M_{\nu}$, and  charged leptons, $M_{\ell}$,  mass matrices which correspond 
to the residual symmetries  $G_\nu$ and  $G_l$. The  
invariance 
means that 
\be
S^T_\nu M_{\nu } S_\nu = M_\nu \, , 
\label{m-inv-cond1}
\ee 
\be
T M_{\ell} M_{\ell}^{\dagger} T^{\dagger} = M_{\ell}M_{\ell}^{\dagger} . 
\label{cl-inv-condition}
\ee
The diagonalization of $M_{\nu}$  and $M_{\ell}$
gives mixing matrix which reflects symmetries $G_\nu$ and $G_l$. 
In fact,  $S_\nu$ and $T$ can be used as the generating elements of  $G_\nu$ and $G_l$. 

The invariance conditions (\ref{m-inv-cond1}) and 
(\ref{cl-inv-condition}) should be satisfied in the basis 
of neutrino and charged lepton states in which the charged current (CC) interactions are  
diagonal. Only in this basis all the information about mixing is contained in the mass matrices. 
This basis is not unique: it is determined upto 
equal rotations of the left handed components of neutrinos and charged 
leptons. (The flavor basis is one possibility.)  
Only in this  basis one can obtain information about mixing and masses 
exploring the mass matrices only~\footnote{Obviously, in the neutrino 
and charged lepton mass bases all the information on mixing 
is in the charged current interactions, and the diagonal mass matrices 
do not contain information about mixing.}.  

Most of the symmetry bounds on lepton mixing in the framework 
(\ref{framework}, \ref{m-inv-cond1}, \ref{cl-inv-condition}) have been obtained via 
{\it explicit model building}. That is, via construction 
of the Lagrangian invariant under $G_f$, spontaneous breaking 
of this  symmetry, and  diagonalization 
of the generated mass matrices. However, consequences of symmetry for mixing (as well as masses) 
can be obtained  {\it without explicit model building} immediately 
from (\ref{framework}, \ref{m-inv-cond1}, \ref{cl-inv-condition}),  
{\it i.e.},  from the conditions that 
 
\begin{itemize}

\item 
the mass matrices of neutrinos and charged leptons 
(in the basis where the CC interactions are diagonal) 
have certain ``residual'' symmetries,  and 

\item 
that these symmetries originate from breaking of the 
original flavor symmetry (or inversely, that they can 
be embedded into this flavor symmetry). 

\end{itemize}
In other words, the relations between the mixing matrix elements 
can be obtained  from the fact that the neutrino and charged lepton 
mass matrices are invariant under transformations 
which generate the flavor symmetry group. 
Relations obtained in specific models must coincide 
with relations obtained here, once $G_\nu$ and $G_l$ 
and the covering group $G_f$ are the same.

\subsection{Symmetry group condition.}
 
The transformations $S_{\nu}$ and $T$ 
in the CC diagonal basis encode information about mixing 
(and masses) that originate from the flavor symmetry.  
The relations between mixing matrix elements
follow from the conditions that $S_{\nu}$ and $T$ 
belong to the same {\it finite discrete} group $G_f$~\cite{dani1, dani2}. 
This means that the product of $S_{\nu}$ and $T$ 
\be
W  \equiv  S_{\nu} \cdot T \,
\label{rel2}
\ee
also belongs to $G_f$. Furthermore, since $G_f$ is a {\it finite} group,
there must exist  integers $n$, $m$ and  $p$ such that 
$S_{\nu}^n = {\bf I}$, $T^m = {\bf I}$ and 
\be
W^p = (S_{\nu} \cdot T)^p = {\bf I}.
\label{wrelation}
\ee
We will call this the  {\it symmetry group condition}.  
The  relations
\be
S_{\nu}^n = T^m = W^p = {\bf I}
\label{rel1}
\ee
form a presentation of $G_f$ which  corresponds to 
the von Dyck group $D(n,m,p)$.
The condition for the group $D(n,m,p)$ to be finite is
\be
\frac{1}{n} + \frac{1}{m} + \frac{1}{p} > 1 \, , 
\label{finitecond}
\ee
and complete list of the finite von Dyck groups 
which have irreducible representation  ${\bf 3}$  
includes 
${\bf D}_n  = D(2,2,n)$,  ${\bf A}_4 = D(2,3,3)$,   
${\bf S}_4 = D(2,3,4)$ and  ${\bf A}_5 = D(2,3,5)$. 

For definiteness,  let us consider the problem in 
the {\it flavor basis},  
which means that the charged lepton mass matrix is diagonal 
and we will use the same notation for 
its symmetry transformations $T$  as before. 
We denote by $S_{\nu U}$ the transformation matrix 
that lefts invariant the neutrino mass matrix   
in the flavor basis:   
\be
S_{\nu U}^T M_{\nu U} S_{\nu U} = M_{\nu U} .   
\label{inflav}
\ee
(The same result can be obtained in any basis where 
the CC interactions are diagonal). 

It is the condition (\ref{wrelation}),  
that connects symmetry transformations of the neutrinos 
and charged leptons, that leads to relations between the 
mixing matrix elements. To see this we  
introduce $S_m$ - the symmetry 
transformation of the neutrino mass matrix   
in the neutrino mass basis:  
\be
S^T_m m_{\nu} S_m = m_\nu \, , ~~~~~ 
m_{\nu} \equiv {\rm diag}\{m_1, m_2, m_3 \} . 
\label{invmass}
\ee
Let us  express $S_{\nu U}$ in terms of $S_m$. 
In the flavor basis the neutrino  mass matrix can be 
represented as 
\be
M_{\nu U} = U_{PMNS}^* m_{\nu } U^\dagger_{PMNS}\, , 
\label{numass}
\ee
where  $U_{PMNS}$ is the PMNS lepton mixing matrix. 
Using Eqs. (\ref{inflav}), (\ref{invmass}) and (\ref{numass}) 
it is easy to show that the  transformation matrix  $S_{\nu U}$ equals
\be
S_{\nu U} = U_{PMNS}S_m U_{PMNS}^\dagger  \,. 
\label{S-flavorbasis}
\ee
Since $S_m$ does not depend on mixing,  
whole information on the mixing is explicit in 
(\ref{S-flavorbasis}). Inserting $S_{\nu} = S_{\nu U}$ into  
(\ref{wrelation}) we obtain 
\be
[U_{PMNS} S_m U_{PMNS}^\dagger T]^p = {\bf I}. 
\label{main5}
\ee
This is the main relation which connects the mixing matrix 
and  generating elements of the group in the mass basis. 
It should give relations between the mixing matrix elements 
in terms of the group parameters. 
Let us stress  that restrictions on mixing  follow from 
the symmetry group condition which includes both 
the neutrino and charged lepton transformations. 

One can obtain the relations for the matrix elements 
immediately from the Eq.  (\ref{main5}), 
however simpler way to solve it   and to analyze solutions 
is the following. It can be shown \cite{dani1} that solution 
of (\ref{main5}) is equivalent to the solution of equation  
\be
{\rm Tr} \left[U_{PMNS}S_m U_{PMNS}^\dagger T\right]  = a \, ,  
\label{witha}
\ee
where $a$ is the sum of  three $p$-th roots of unity, $\lambda_\alpha$: 
\be
a = \sum_{\alpha = 1}^3 \lambda_\alpha, ~~~~  (\lambda_\alpha)^p = 1 . 
\ee
The $p$-th roots of unity are a finite set of complex numbers,
and in general (especially for large $p$) several complex 
values for $a$  can be found~\footnote{Here we use the same definition 
of $a$ as in the paper \cite{dani2} which has an opposite sign 
with respect to definition given in \cite{dani1}.}.  
In fact, $\lambda_\alpha$ are the eigenvalues of $W$ 
they can be parameterized \cite{lam13} as 
\be
\lambda_\alpha = e^{i \xi_\alpha}, ~~~~ 
\xi_\alpha = 2\pi \frac{s_\alpha}{p},~~~~ (\alpha = e, \mu, \tau),  
\ee
where integers $s_\alpha$ satisfy inequality $s_\alpha < p$. 
Since ${\rm Det} [W] =  \lambda_e \lambda_\mu \lambda_\tau =   1$, 
we have $s_\tau = - s_e - s_\mu$, and consequently, 
$a = a (p, s_e, s_\mu)$. 
For known $a$ and given $S_m$ and $T$, Eq.~(\ref{witha}) 
provides a complex condition that the entries of $U_{PMNS}$ must satisfy.

Notice that instead of  (\ref{rel2}) one can consider some other products of $S_\nu$ and $T$  
or even two products simultaneously.  The latter should not lead 
though to new relations between the mixing elements but will impose 
constrains on the group parameters (see below).  

The symmetries $G_\nu$ and $G_l$ are broken in whole 
the theory, and therefore one expects corrections to the results 
obtained in the symmetry limit. Corrections can not be taken into account 
in this model independent formalism. If however, the residual symmetry 
of mass matrix exists (or imposed) after the corrections are taken into account, then 
the relations exist for mixing parameters with corrections.

\subsection{Generic invariance of mass matrices.}

There are certain symmetries of  $M_\nu$ and $M_{\ell}$ 
which are always present. They play important role in our consideration.  
For arbitrary values of masses the neutrino mass matrix 
$m_{\nu} \equiv {\rm diag}\{m_1, m_2, m_3 \}$
is invariant under the transformations 
\be
S_1 = {\rm diag}\{1,\,-1,\,-1\}\,,\quad S_2 = {\rm diag}\{-1,\,1,\,-1\}\,,\quad
\label{Si}
\ee
and  $S_3 = S_1 S_2$ with   ${\rm Det}[S_i] = 1$:  
\be
S^T_i m_{\nu } S_i = m_\nu \, .  
\label{m-inv-cond} 
\ee
The transformations $S_{1}$ and $S_{2}$ 
satisfy  conditions $S_{i}^2  = {\bf I}$ and  
generate the  Klein group ${\bf Z}_2\otimes {\bf Z}_2$. 
Then in the flavor basis  the mass matrix $M_{\nu U}$ is invariant 
under the transformation $S_{iU} =  U_{PMNS} S_i U_{PMNS}^{\dagger}$. 

The charged lepton mass matrix in the flavor basis, 
$m_{\ell} = {\rm diag}(m_e,~ m_\mu, ~ m_\tau)$,  
has a full $[U(1)]^3$ symmetry. 
We assume that the residual discrete symmetry  is a 
${\bf Z}_m$ subgroup of $[U(1)]^3$. 
So, the corresponding transformation  
which satisfies 
the condition $T^m = {\bf I}$, can be written as
\be  
T \equiv {\rm diag} \{e^{i\phi_e},\, e^{i\phi_\mu},\, e^{i\phi_\tau}\},  
\label{Tdef}
\ee
where 
\be
\phi_\alpha \equiv 2\pi \frac{\kappa_\alpha}{m} \,, \quad   \alpha = e,\, \mu,\, \tau \; ,  
\label{phidef}
\ee
with $\kappa_\alpha \leq m$. 
Since the flavor group is a subgroup of $SU(3)$, we have  
${\rm Det}[T] = 1$, which gives 
\be
\phi_e + \phi_\mu + \phi_\tau = 0,
\ee
or equivalently, $\kappa_\tau = - \kappa_e + \kappa_\mu$. 

Notice that $T$ and $W$ enter the formalism in similar way  
and can be interchanged with substitution 
$m \leftrightarrow p$, $\kappa_\alpha \leftrightarrow s_\alpha$.

Recall that  symmetries generated by $S_{iU}$ and $T$ are always 
present and they can not be broken. The invariance does 
not depend on specific values of masses. Therefore the generic symmetries
lead to relations between mixing parameters without connection to masses. 
In particular, it is for this reason the generic symmetry based on 
${\bf Z}_2\otimes {\bf Z}_2$ was used in models of the TBM mixing. 
The symmetry transformations are present, but may or may not 
form the flavor group $G_f$, that is, they may or may not satisfy the 
symmetry group condition (\ref{wrelation}).

\section{Mixing patterns for different $G_\nu$ }

In what follows we will use $T$ as in (\ref{Tdef}), and  
consider different $G_\nu$. 

\subsection{$G_\nu = {\bf Z}_2$}

Let us take one  ${\bf Z}_2$ as $G_\nu$, {\it i.e.} only one generic 
(``mass independent'') symmetry of the neutrino mass matrix.    
This means that only one transformation, e.g.,  $S_{1U}$
satisfies the group condition 
$(S_{1U} \cdot T)^p = 1$,  whereas two others - don't:   
$(S_{iU} \cdot T)^q \neq 1$, $i = 2, 3$  for any $q$. 
They are outside of the group and do not lead to bounds on the mixing 
matrix.  Obviously, $S_{i}^2 = {\bf I}$ and therefore $n = 2$ in 
the presentation (\ref{rel1}).  

With $T$ from (\ref{Tdef}) the condition  
(\ref{witha}) can be written  explicitly as   
\be
\sum_\alpha e^{i\phi_{\alpha}} (2|U_{\alpha i}|^2 -1)= a,   
\label{condaexp}
\ee
or introducing $a \equiv a_R + i a_I$, as
\be
\sum_\alpha  \cos \phi_{\alpha} (2|U_{\alpha i}|^2 -1) = 
a_R, ~~~~~ 
\sum_\alpha  \sin \phi_{\alpha} (2|U_{\alpha i}|^2 -1) = 
a_I.   
\label{condaexp2}
\ee
These relations

\begin{itemize}

\item
depend on the {\it moduli} of matrix elements, which is related to 
the diagonal form of the generating elements $S_j$ and $T$;  

\item
relate  elements of a {\it single column} $j$,    
and  this index  $j$ is determined 
by the index $j$ of the neutrino transformation matrix $S_j$;     

\item
determine the column $j$ {\it completely}, 
since two relations (for real and imaginary parts of $a$)
plus unitarity are imposed. 

\end{itemize}

Explicitly the relations read: 
\begin{eqnarray}
|U_{ej}|^2 & = & \frac{a_R \cos \frac{\phi_e}{2} + 
\cos\frac{3\phi_e}{2} - a_I\sin \frac{\phi_e}{2} }{4\sin{\frac{\phi_{e\mu}}{2}} 
\sin{\frac{\phi_{\tau e}}{2}}}\,, 
\label{Uei} 
\nonumber \\
|U_{\mu j}|^2 & = & \frac{a_R \cos \frac{\phi_\mu}{2} + \cos\frac{3\phi_\mu}{2} 
- a_I\sin \frac{\phi_\mu}{2} }{4\sin{\frac{\phi_{e\mu}}{2}} \sin{ \frac{\phi_{\mu\tau}}{2}}} \,,\label{Umui} 
\nonumber\\
|U_{\tau j}|^2 & = & \frac{a_R \cos{\frac{\phi_\tau}{2}} 
+ \cos{ \frac{3\phi_\tau}{2}} - 
a_I\sin{ \frac{\phi_\tau}{2}}}{4\sin{\frac{\phi_{\tau e}}{2}} \sin{ \frac{\phi_{\mu\tau}}{2}}}\,, 
\label{Utaui} 
\end{eqnarray}
where 
\be
\phi_{\alpha \beta} \equiv \phi_\alpha - \phi_\beta \,,\quad \alpha,\,\beta = e\,,\mu,\,\tau \, .  
\ee
Due to the unitarity,  
$\sum_{\alpha}|U_{\alpha j}|^2 = 1$,  there are two independent relation.  
For specific $S_j$, the equations give the absolute values 
of the $j$-th column of $U_{PMNS}$,  
if the values of $m$, $p$, $\kappa_e$, $\kappa_\mu$ and $a$ are given. 
In turn,  $a = a (p, s_e, s_\mu)$, so that mixing elements are determined 
by 6 integer parameters (here $n = 2$) which fix eigenvalues of 
$T$ and $W$: ($m,~\kappa_e,~ \kappa_\mu,~ p, ~ s_e,~ s_\mu$). 
A choice of these parameters is, however, restricted  
by the fact that $T$ and $W$ form a finite group, 
see Sec. 4.2  and \cite{lam13}.  
Substituting the standard parametrization for $U_{PMNS}$ 
in Eqs.~(\ref{Utaui}) one obtains the two conditions that the mixing angles 
and the CP phase must satisfy. 

If one of the charged leptons has the $T$-charge zero,  
$\kappa_\alpha = 0$ (and  so two others have opposite signs),  
the expressions (\ref{Utaui}) simplify: 
\begin{eqnarray}
|U_{\alpha j}|^2 & = & \eta, ~~~~  
\label{main2} \\
|U_{\beta j}|^2 &  = & |U_{\gamma j}|^2 = \frac{1 - \eta}{2} \, , ~~~~~ \beta,\,\gamma \neq \alpha, 
\label{main1}
\end{eqnarray}
where 
\be
\eta  \equiv \frac{1 + a}{4 \sin^2\left( \frac{\pi k}{m} \right)}. 
\label{eta}
\ee
Recall that here $\alpha$ corresponds to the charged lepton which is invariant under $T$ transformation, 
$j$ corresponds to  $S_j$. All the mixing parameters are determined by 
single quantity $\eta$ which, in turn, is the function of the group parameters $a(p)$, $m$, $\kappa_e$, 
and $\kappa_\mu$. In other words $\eta$ is determined by the group assignment: 
$\{p, m, \kappa_e, \kappa_\mu, a \}$. For small $p$, 
the parameter $a$ is determined by $p$ uniquely.  
 
For finite groups we have found from (\ref{Utaui}) and 
(\ref{main2} - \ref{eta}) the following 
phenomenologically interesting possibilities 
\be
\frac{1}{6}\left(
\begin{array}{l}
 4 \\
 1  \\
 1
\end{array}
\right), ~~~~~
\frac{1}{3} \left(
\begin{array}{l}
 1 \\
 1  \\
 1
\end{array}
\right),  ~~~~~
\frac{1}{4}\left(
\begin{array}{l}
 1 \\
 2  \\
 1
\end{array}
\right). 
\label{TBM12}
\ee
The first possibility corresponds to 
$\{p, m, \kappa_e, \kappa_\mu, a\}  = \{4, 3, 0, 1, 1\}$ and is called 
the trimaximal mixing-1.  
It can be used as the first column ($j = 1$) of the mixing matrix. 
The second one called the trimaximal mixing-2 
can be obtained  for $\{p, m, \kappa_e, \kappa_\mu, a) = (3, 3, 0, 1, 0\}$.  
It is viable solution for $j = 2$. 
Also the last possibility realized for  $\{p, m, \kappa_e, \kappa_\mu, a\} = \{3, 4, 3, 0, 1\}$ 
with certain corrections can be used for  
$j = 2$.  Another class of possibilities can be obtained by 
($m \leftrightarrow p$) permutation which corresponds to the exchange  
$T \leftrightarrow W$ in a group presentation \cite{dani2}. 

Notice that the first as well as the second columns 
in (\ref{TBM12}) coincide with columns of 
the TBM mixing. This means that TBM is  special point in the parameter space 
of the solutions determined by one of the columns.

\begin{figure}[h]
\begin{center}
\includegraphics[width=17pc]{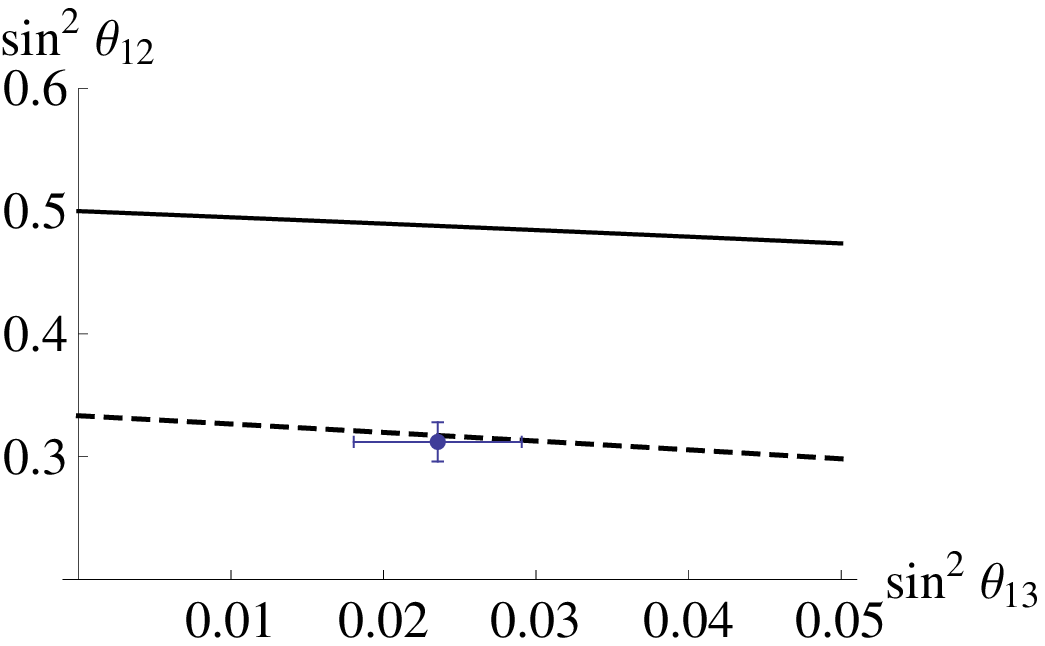}\hspace{2pc}%
\includegraphics[width=17pc]{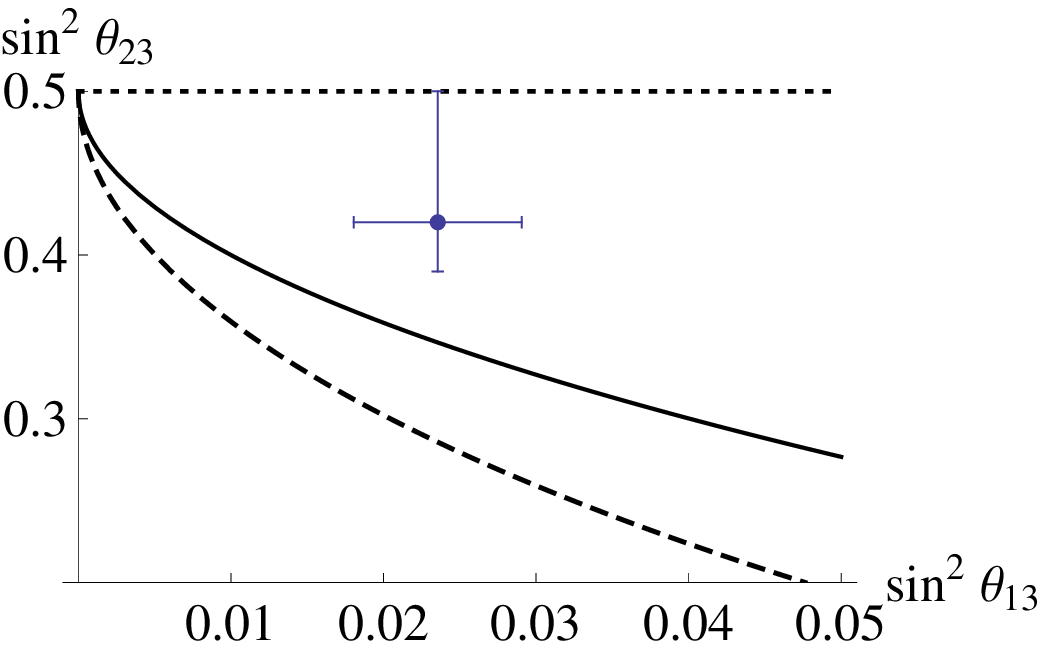}\hspace{2pc}%
\end{center}
\caption{\label{fig:s4}
Symmetry relations between the mixing angles in the case of $G_f = {\bf S}_4$ and 
$G_\nu = {\bf Z_2}$. 
Shown are dependences of $\sin^2 \theta_{12}$ on $\sin^2 \theta_{13}$ (left panel), 
and $\sin^2 \theta_{23}$ on $\sin^2 \theta_{13}$ for $\delta = \pi$ 
(right pannel) for two different symmetry assignment: 
$\{p, m, \kappa_e, \kappa_\mu, a\} =  \{4, 3, 0, 1, 1\}$  and  $j = 1$ (dashed lines) and 
$\{p, m, \kappa_e, \kappa_\mu, a\} = \{3, 4, 0, 3, 0\}$ (solid lines). 
The dotted line in the right panel corresponds to 
$\delta = \pi/2$ for both symmetry assignments. 
Crosses show the experimental results. 
}
\end{figure}

Using the standard parameterization  for the PMNS matrix one can 
find from (\ref{main2} - \ref{eta})  
relations between the mixing angles and the CP - phase. 
Eqs. (\ref{Utaui}) and (\ref{main2}) impose two relations 
on 4 parameters. Therefore using two known mixing angles one can predicts the CP-phase and 
the  angle which is not yet well known (e.g., $\theta_{23}$).  

As an example, we show in Fig.~\ref{fig:s4} the relations for 
two different symmetry assignments. 
The assignment $\{p, m, \kappa_e, \kappa_\mu, a\} =  \{4, 3, 0, 1, 1\}$  with  $j = 1$ 
(it corresponds to $G_f = {\bf S}_4$ and 
trimaximal-1) gives dependences which agree very well with 
the experimental data. Furthermore, according to the right panel of Fig. 1, one expects 
$\delta = 90^{\circ} - 120^{\circ}$.


\subsection{Finite subgroups of the infinite von Dyck groups.}

Number of possibilities which lead to finite groups 
is very restricted. It can be extended if we consider big ($>5$) numbers  
$m, n, p$. This, however,  
makes group infinite, and therefore in general condition  
(\ref{rel1}) can not be imposed. 
In order to render the group finite, an additional relation between 
the generating elements $S_{iU}$ and $T$ 
should be added to Eq.~(\ref{rel1}). 
One possibility is the relation of the form
\be
X^q  = 1\,,\quad X \equiv S_{jU}T^{-1}S_{jU}T   
\label{relX} \, 
\ee 
or explicitly
\be
(S_{iU}T^{-1}S_{iU}T)^q = 1 . 
\ee
Together with (\ref{rel1}), that would correspond to certain 
modular groups: $PSL(2,\, \mbf{Z}_7)$ 
if  $\{p,\,m\} = \{7, 3\}$ or $\{7,\,4\}$, 
$\Delta(96)$ if $\{p,\,m\} = \{8, 3\}$  and   
$\Delta(384)$ for  $\{p,\,m\} = \{16, 3\}$ \cite{toorop}.

Using the same procedure as before for 
the matrix $W$,  we obtain that Eq.~(\ref{relX})  
leads to the condition  
\be
{\rm Tr}[X] = x,   
\label{condx}
\ee
where
\be
x = \sum_\beta \lambda_\beta^{\prime}, ~~~~ \lambda_\beta^{\prime q} = 1. 
\label{eqx}
\ee
The equation for $X$ gives relations between the 
mixing elements of the same column $j$ (since the same $S_j$ is involved) which have 
already been completely 
fixed by the condition for $W$. The relations should be consistent. 
Therefore the condition Eq.~(\ref{relX}) does not add 
new constraints on mixing angles, but imposes constrains 
on the group parameters to be consistent with 
conditions from $W$.  Indeed,  
${\rm Tr} [S_{iU}T^{-1}S_{iU}T]$ (and therefore $x$) 
can be expressed in terms of ${\rm Tr}[W] = a$ 
and other group parameters as~\cite{dani2} 
\be
x = |a|^2 + a{\rm Tr}[T^\dagger] + a^\dagger {\rm Tr}[T] + A , 
\label{conda'exp}  
\ee
where 
\be
A = 2 + e^{i\phi_{e\mu}} + e^{-i\phi_{e\mu}} + 
e^{i\phi_{\mu\tau}} + e^{-i\phi_{\mu\tau}} + 
e^{i\phi_{\tau e}} + e^{-i\phi_{\tau e}}\, .
\ee
The equation (\ref{conda'exp}) 
should be considered as condition for $x$, $a$  and phases 
$\phi_\alpha = \phi_\alpha (m, k_\alpha)$.  
So,  it is essentially a condition for 
the matrix $T$ that fixes the values of $k_e$ and $k_\mu$ and  
makes the group finite. 
Instead of $k_\alpha$ one can impose bounds on $a$ or  $s_\alpha$.  
(See discussion in \cite{lam13} where 
the bounds on these parameters have been obtained by scanning of the groups.)  
Imposing the second condition gives systematic way to get bounds on the group parameters  

Let us consider two examples~\cite{dani2}. 

\begin{figure}[h]
\begin{center}
\includegraphics[width=17pc]{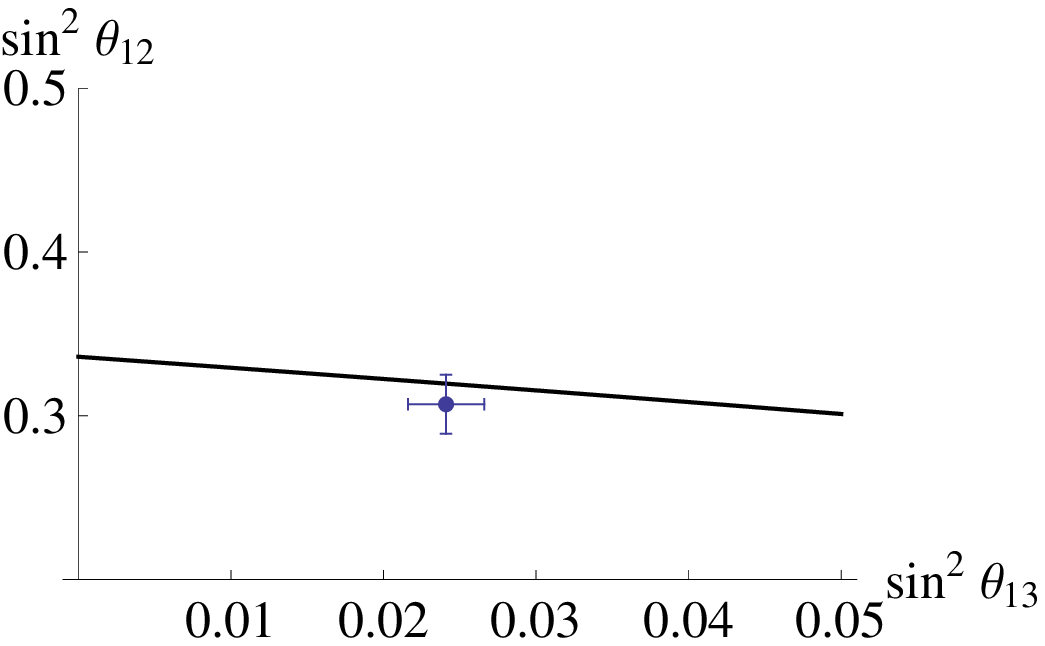}\hspace{2pc}%
\includegraphics[width=17pc]{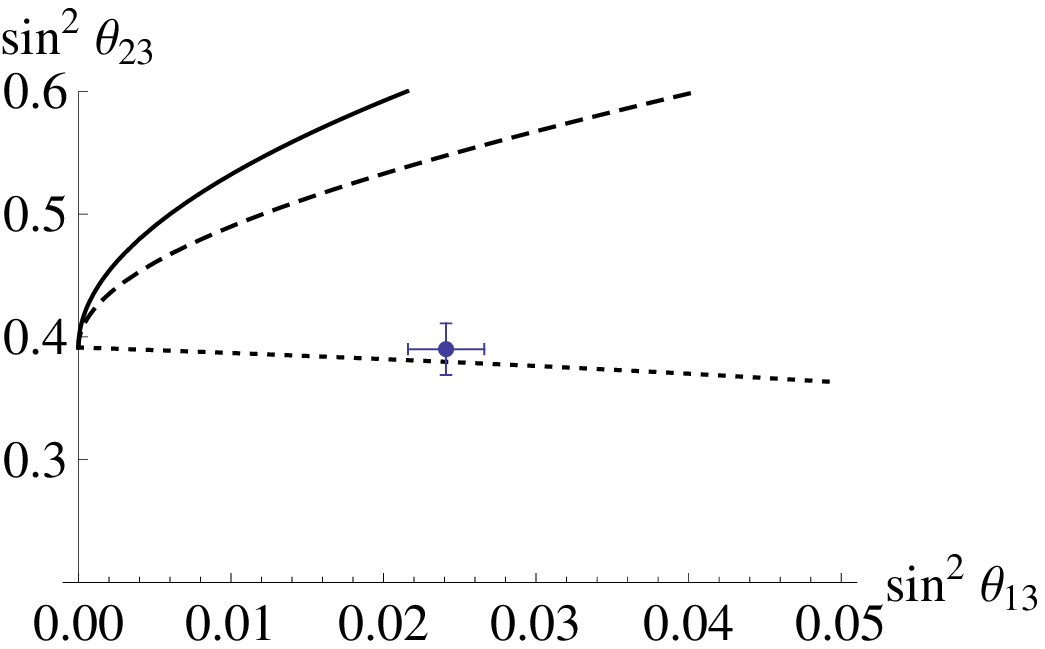}\hspace{2pc}%
\end{center}
\caption{\label{fig:psl}
Symmetry relations between mixing angles in the case of $G_f = 
PSL(2, {\bf Z}_7)$ and  $G_\nu = {\bf Z_2}$.
{\it Left panel:} dependence of $\sin^2 \theta_{12}$ on $\sin^2 \theta_{13}$, {\it right panel:} 
dependence of $\sin^2 \theta_{23}$ on $\sin^2 \theta_{13}$ for 
$\delta = 0$ (solid line),   
$\delta = \pi/4$ (dashed line)  
and $\delta = \pi/2$ (dotted lines). The symmetry assignment is
$\{p, m, \kappa_e, \kappa_\mu, a \} =  \{3, 7, 5, 3, 0\}$ 
and  $j = 2$. Crosses show the experimental results.}
\end{figure}

1).  The group $PSL(2,\,\mbf{Z}_7)$ is a subgroup of the infinite von Dyck group with 
the presentation (see Eq.~(\ref{rel1}))
\be
S_{iU}^2 = T^7 = (S_{iU}T)^3 = {\bf I} \, 
\ee
({\it i.e.}, $m=7$)  and with an  additional condition  
\be
X^4 = {\bf I} \label{X^4}\, 
\ee
(Eq.~(\ref{relX}) with $q=4$). 
For symmetry assignment 
$
\{p, m, \kappa_{e}, \kappa_\mu  , a \}  = \{3, 7, 5, 3, 0 \},  
$
we  obtain $x = 0$  and  values of the mixing elements
\be
|U_{\mu j}|^2 = \frac{1}{4\left[ 1 + \sin \frac{\pi}{14} \right]} \,,\quad |U_{\tau j}|^2 = 
\frac{1}{4 \left[ 1 + \cos \frac{\pi}{7} \right]} , 
\label{Umu1PSL7} 
\ee
and 
$|U_{ej}|^2 = 1 - |U_{\mu j}|^2 -  |U_{\tau j}|^2$. 
For $j = 2$ this gives good description of the experimental data.  
The corresponding relations for mixing angles are shown in  
 Fig. \ref{fig:psl}. According to the right panel within 1$\sigma$ allowed region 
of mixing angles the CP phase equals $\delta = (80^\circ - 90^\circ)$. 

2). The group $\Delta(384)$ is a subgroup of the infinite 
von Dyck group with presentation
\be
S_{iU}^2 = T^{16} = (S_{iU}T)^3 = {\bf I} \,,
\ee
and  additional relation
\be
X^3 = {\bf I} .  
\label{rel6}
\ee
Again, we have  $x = 0$.
Viable mixing pattern can be obtained in the case of $W \leftrightarrow T$ 
permuted version, when $W$ is the symmetry of the charged leptons and 
$T = S_{iU} W$ gives the symmetry group condition. Now $p^{\prime} = m$, $m^{\prime} = p$, 
$a^{\prime} = {\rm Tr} [T]$ and $\kappa_\alpha \rightarrow \xi_\alpha$. 
For the set of the group parameters
\be
\{p^{\prime},  m^{\prime}, 
s_e, s_\mu , a^{\prime}  \} =  \left\{16,  3,  1,  1, \frac{- 1+i}{\sqrt{2}}
\right\} \, 
\ee
we obtain 
\be
\{|U_{ej}|^2,\, |U_{\mu j}|^2,\, |U_{\tau j}|^2\} 
= \left\{ \frac{4+\sqrt{2}+\sqrt{6}}{12},\, \frac{4+\sqrt{2}-\sqrt{6}}{12},\, 
\frac{2-\sqrt{2}}{6} \right\} . 
\ee
If  $j = 3$,  this model is able to fit all mixing angles for 
$\delta \sim 50^{\circ} - 60^{\circ}$.

\subsection{$G_\nu = {\bf Z}_2\otimes {\bf Z}_2 $.}

Let us consider the  complete generic symmetry given by the  Klein group  
$G_\nu = {\bf Z}_2\otimes {\bf Z}_2 $.  
A presentation of  the group is given (for definiteness we take $S_1$, $S_2$ as generating elements) by
\be
S_{1U}^2 = S_{2U}^2 = T^m = W_{1U}^{p_1} = W_{2U}^{p_2} = {\bf I} \,,\quad [S_{1U},\,S_{2U}] = 0.
\label{kleing-rels}
\ee
They impose two sets of conditions on $U_{PMNS}$ which can be written as
\be
{\rm Tr}{[W_{jU}]} = \sum_{\alpha} e^{i\phi_\alpha} 
(2|U_{\alpha j}|^2 -1)  
 = a_j \,,\quad ~~~ j = 1,\, 2,   
\label{ai}
\ee
where $a_j \equiv \sum_k \lambda^{(j)}_k$ are sums 
of three $p_j$-th roots of unity. Now there are four 
relations between the mixing matrix elements, corresponding to two columns of the mixing matrix, 
which  determine $U_{PMNS}$ completely~\cite{dani2}. 
An example of this case is given by two columns in Eq.~(\ref{TBM12})  
with  $\{p_1, \,p_2, \, m \} = \{4,\,3, \, 3 \}$. 
The only mixing matrix compatible with 
first two columns in (\ref{TBM12}) is the TBM.
In a sense the TBM is special being related to maximal 
generic neutrino symmetry  group.

Complete scan of the discrete finite groups with  
order less than 1536 and 
$G_\nu = {\bf Z}_2\otimes {\bf Z}_2 $  
has been performed in~\cite{manfred}
and for these groups the mixing pattens has been 
computed~\footnote{The approach in \cite{manfred} is, however,  different 
from the one presented here:
the starting point in~\cite{manfred}  is the explicit form of the generators 
of the residual symmetries for leptons and neutrinos in 
the basis of diagonal CC interaction. The generators satisfy 
presentation of the full selected flavor group.  
The generators are diagonalized by rotations 
$\Omega_\ell$ and $\Omega_\nu$. Then, as can be shown, 
the mixing matrix  is given by $\Omega_\ell^{\dagger} \Omega_\nu$.  
See also discussion in \cite{lam13}.}.\\

The formalism allows to explain an observation 
\cite{toorop} that in some cases, the mixing matrix derived from Eqs.~(\ref{ai}) 
has the property that the absolute values of  entries 
of two column were equal up to a permutation: 
\be
|U_{\alpha i}|^2 = |U_{f(\alpha) j}|^2,  ~~~i \neq j .
\ee 
Here $f(\alpha)$ is the permutation operation of the 
flavor indices. An example, which is realized in  
the case of $PSL(2,\,\mbf{Z}_7)$  group is 
\be
|U_{\alpha i}|^2  = \left( \begin{array}{ccc}
c_1 & c_2 & c_3 \\
c_3 & c_1 & c_2  \\
c_2 & c_3 & c_1
\end{array} \right) \,  \label{PSL7fullU}
\ee
with
\be
c_2 = \frac{1}{4\big( 1 + \sin\frac{\pi}{14}\big)} \,,\; ~~~~c_3 
= \frac{1}{4\big(1 + \cos\frac{\pi}{7}\big)} \, 
~~~~~
c_1 = 1 - c_2 - c_3.  
\ee

\subsection{Non-generic neutrino symmetry.}

We can introduce a non-generic  neutrino 
symmetry $G_\nu$ which includes  transformation, $Q_m$,  
under which  neutrino  mass matrix in the mass basis 
is invariant only for some specific values of masses.   
Inversely, introduction of such a symmetry in the flavor symmetry 
imposes conditions on neutrino masses. 
The symmetry group condition for $Q_m$ reads 
$(Q_{mU}\cdot T)^r = 1$, 
where $r$ is integer and  $Q_{mU} = U_{PMNS} Q_m U_{PMNS}^{\dagger}$ 
is the symmetry transformation in the flavor basis.  
Also  some of generic symmetries ${\bf Z}_2$ could be a part of $G_f$. 
In the latter case one should impose also the relation  
$(Q_{mU} \cdot S_U)^q = (Q_m \cdot S)^q =  1$. 
The mixing matrix disappears in the last equality   
and therefore it should be considered as a consistency condition. 

\section{Conclusion}

1. The appealing discrete flavor symmetry framework
(inspired by TBM)  is based on idea that the lepton mixing originates 
from different ways of the discrete flavor symmetry breaking in the 
neutrino and charged lepton Yukawa sectors. 
These different ways lead to different residual symmetries 
of the neutrino and charged lepton mass matrices 
which ensure special forms of these matrices, and consequently,  
special form of the mixing matrix. Here the 
symmetry transformation are generic transformations 
valid for arbitrary masses.

2. Recent measurements of the neutrino oscillation parameters 
show substantial  deviations from the TBM mixing. This may further 
indicate that TBM is accidental and whole discrete 
flavor symmetry approach is phenomenologically irrelevant. 
Experimental value of the 1-3 mixing can be 
connected with other observables in various ways 
which have different implications to theory. 
It is not clear  however if these relations imply discrete symmetries: 
some -- certainly not, although symmetry effects can be 
rather hidden. 

Further developments in the field can be related to 
establishing the neutrino  mass hierarchy. 
``Race'' for the mass hierarchy and CP has started:  
studies of the atmospheric neutrinos with multi-megaton mass detectors having 
low energy thresholds can provide fast, inexpensive and reliable answer.  

3. Discrete symmetries still  may play important role in 
formation of the lepton flavor structures. 
TBM can be  treated  as the lowest order structure which requires  corrections. 
Discrete symmetries can be consistent 
with the non-zero 1-3 mixing and deviation of the 2-3 mixing 
from maximal. 

4. Consequences of symmetries for mixing, and in general, for masses 
in this framework can be obtained immediately 
from the {\it symmetry group condition(s)}, without explicit model building. 
For this  the knowledge of (assumption about)
the residual symmetries and the covering group is  enough.  
The symmetry group condition includes the PMNS matrix and the 
generating elements of the residual symmetries in the mass basis. 

From the point of view of specific models, here it is assumed that 
the relevant model-building has already been done and the 
mass matrices of neutrinos and charged leptons with 
certain symmetries obtained.

5. The symmetry group conditions have been  applied  to  
different residual neutrino symmetries and full flavor groups. 
(i) The  generic $G_\nu = {\bf Z}_2$ with  finite von Dyck group 
as well as  finite subgroup of the infinite von Dyck group  as $G_f$ lead to 
two relations between the mixing parameters. This   
fixes one of the columns of mixing matrix. 
(ii) The  Klein group   ${\bf Z}_2 \times {\bf Z}_2$ as $G_\nu$ 
fixes the mixing matrix completely.    
(iii) The formalism can be generalized to 
include non-generic  neutrino symmetries $G_\nu$ which lead to  
relations between mixings and neutrino masses.  

6. The relations between mixing parameters obtained here 
include the CP-phase and therefore  the  phase can be predicted 
using measured values of mixing angles. In the examples we 
discussed  $\delta$ can be  around $60^{\circ}$ or $90^{\circ}$. 
So, future measurements of the phase can provide 
important test of the framework.

7. The formalism  presented here can be used in various ways.  
In specific model, once the residual symmetries are identified 
the consequences on mixing and masses can be obtained immediately. 
The formalism allows to analyze systematically  possible 
mixing patterns that can be extracted from the finite 
groups or subgroups of the von Dyck type. 
Inversely, for a given pattern of mixing the formalism 
allows to perform the ``symmetry building'' - identify 
the residual and covering symmetry groups. 

The formalism allows also to 
explain various features of mixing matrix which follow from symmetries. 
In particular, in the case of Klein group in the neutrino sector 
it can explain appearance of mixing patterns in which two or three columns 
of $U_{PMNS}$ have equal but  permuted elements.

8. Further generalizations of the formalism are possible.  
On the other hand, discrete symmetries can be realized in some other way in   
frameworks which differs from  (\ref{framework}).  
Mixing (or its zero order structure)  can originate from  
different nature of the mass terms of the charged leptons 
(Dirac)  and  neutrinos (Majorana), or from neutrino 
mixing with new degrees of freedom (e.g.,  singlets of SM). 

\subsection{Acknowledgments}
Large part of this talk is based on papers written with  
in collaboration with D. Hernandez.  
I am grateful to C. Hagedorn, A. Pilaftsis and R. Mohapatra  
for useful discussions during the Symposium. 

\section*{References}

\end{document}